\documentclass[sigconf]{acmart}

\acmConference[ICSE 2022]{The 44th International Conference on Software Engineering}{May 21–29, 2022}{Pittsburgh, PA, USA}

\usepackage{balance}
\usepackage{amsmath}
\usepackage{array}
\usepackage{lipsum}
\usepackage{listings}
\definecolor{mGreen}{rgb}{0,0.6,0}
\definecolor{mGray}{rgb}{0.5,0.5,0.5}
\definecolor{mPurple}{rgb}{0.58,0,0.82}
\definecolor{backgroundColour}{rgb}{0.95,0.95,0.92}

\lstdefinestyle{CStyle}{
    backgroundcolor=\color{backgroundColour},   
    commentstyle=\color{mGreen},
    keywordstyle=\color{magenta},
    numberstyle=\tiny\color{mGray},
    stringstyle=\color{mPurple},
    basicstyle=\scriptsize\ttfamily,
    breakatwhitespace=false,         
    breaklines=true,                 
    captionpos=b,                    
    keepspaces=true,                 
    numbers=left,                    
    numbersep=5pt,                  
    showspaces=false,                
    showstringspaces=false,
    showtabs=false,                  
    tabsize=2,
    language=C
}

\usepackage{algorithm}
\usepackage{algpseudocode}

\usepackage{xspace}
\usepackage{blindtext}
\usepackage{hyperref}

\usepackage{multirow}

\newcommand{\APPR}{\emph{HUDD}\xspace}

\newcommand{\IEE}{IEE}

\AtBeginDocument{%
  \providecommand\BibTeX{{%
    \normalfont B\kern-0.5em{\scshape i\kern-0.25em b}\kern-0.8em\TeX}}}

\copyrightyear{2022} 
\acmYear{2022} 
\setcopyright{rightsretained} 
\acmConference[ICSE '22 Companion]{44th International Conference on Software Engineering Companion}{May 21--29, 2022}{Pittsburgh, PA, USA}
\acmBooktitle{44th International Conference on Software Engineering Companion (ICSE '22 Companion), May 21--29, 2022, Pittsburgh, PA, USA}
\acmDOI{10.1145/3510454.3516858}
\acmISBN{978-1-4503-9223-5/22/05}

\begin{document}

\title{HUDD: A tool to debug DNNs for safety analysis}


\author{Hazem Fahmy}
\affiliation{%
 \institution{SnT Centre, University of Luxembourg}
 \streetaddress{JFK 29}
 \city{Luxembourg}
 \country{Luxembourg}}
\email{hazem.fahmy@uni.lu}

\author{Fabrizio Pastore}
\affiliation{%
 \institution{SnT Centre, University of Luxembourg}
 \streetaddress{JFK 29}
 \city{Luxembourg}
 \country{Luxembourg}}
\email{fabrizio.pastore@uni.lu}

\author{Lionel Briand}
\affiliation{%
 \institution{SnT~Centre,~University~of~Luxembourg}
 \streetaddress{JFK 29}
 \city{School of EECS, University of Ottawa}
 \country{Ottawa, Canada}}
\email{lionel.briand@uni.lu}


\begin{abstract}

We present \APPR, a tool that supports safety analysis practices for systems enabled by Deep Neural Networks (DNNs) by automatically identifying the root causes for DNN errors and retraining the DNN.
\APPR stands for Heatmap-based Unsupervised Debugging of DNNs, it automatically clusters error-inducing images whose results are due to common subsets of DNN neurons. 
The intent is for the generated clusters to group error-inducing images having common characteristics, that is, having a common root cause.

\APPR identifies root causes by  applying a clustering algorithm to matrices (i.e., heatmaps) capturing the relevance of every DNN neuron on the DNN outcome. Also, \APPR retrains DNNs with images that are automatically selected based on their relatedness to the identified image clusters. Our empirical evaluation with DNNs from the automotive domain have shown that \APPR automatically identifies all the distinct root causes of DNN errors, thus supporting safety analysis. Also, our retraining approach has shown to be more effective at improving DNN accuracy than existing approaches.

A demo video of \APPR is available at\\ \url{https://youtu.be/drjVakP7jdU}.

\end{abstract}




\maketitle

\section{Introduction}
\label{sec:introduction}

Deep Neural Networks (DNNs) are common building blocks for safety-critical cyber-physical systems (e.g., their perception layer), particularly in the automotive sector. 
Common examples include Advanced Driver Assistance Systems (ADAS), where DNNs are used to automate emergency braking and lane changing~\cite{TeslaDNN}.
DNN-enabled systems are a key product not only for large companies but also for car component manufacturers~\cite{IEE,ZF}. This is the case of \IEE Sensing~\cite{IEE}, our industry partner, who develops in-vehicle monitoring systems such as drowsiness detection and gaze detection systems~\cite{Naqvi2018}.

When DNN-based systems are used in a safety-critical context (e.g., automotive), 
developers must comply with 
safety standards such as ISO26262~\cite{ISO26262} and ISO/PAS 21448~\cite{SOTIF}.
Such safety standards enforce the identification of the situations in which the system might be unsafe (i.e., outputs leading to safety violations) and the design of countermeasures to put in place (e.g., integrating different types of sensors). 
However, since DNNs transform high-dimensional vectors through a large number of activation functions whose parameters are learned during training, 
engineers cannot understand the rationale of predictions through manual inspection of DNN code and, consequently, they cannot rely on traditional safety analysis practices. For this reason, safety standards targeting DNN-enabled systems (e.g., ISO/PAS 21448) suggest (1) to rely on accuracy evaluation (i.e., test the DNN using inputs generated by simulators or collected in the field) to identify unsafe scenarios and (2) to rely on the manual inspection of error-inducing images to perform \emph{root cause analysis} (i.e., to understand what are the characteristics of the inputs that lead to a DNN error). 

In this context, quantitative targets for accuracy evaluation may be used to demonstrate that unsafe situations are unlikely; however, standards like ISO/PAS 21448 point out that quantitative targets are not sufficient and that engineers remain liable for potentially hazardous scenarios underrepresented in the test set.

Root cause analysis is expected to help reduce such liability risk. Indeed, engineers can retrain the DNN with additional images similar to the ones leading to DNN errors; however, \emph{the manual identification of additional inputs to retrain the DNN is expensive}.
Also, engineers can introduce countermeasures to make the system robust against unsafe conditions (e.g., by relying on both radar and vision to take decisions).
Unfortunately, \emph{the manual identification of unsafe conditions is error-prone}. For example, engineers may overlook unsafe conditions that are underrepresented in the test set. 
Indeed, human body simulators may generate head images with an horizontal angle determined based on a uniform distribution, between 160 (head turned right) and 220 degrees (head turned left). As a result, very few images with an angle of 160 or 220 degrees are generated and, though it may be an unsafe condition (i.e., one eye is barely visible and the gaze direction prediction may be inaccurate) experiments based on test sets generated with such simulators may suggest that the DNN is on average very accurate and engineers may not notice such unsafe conditions. 


Existing toolsets do not help engineers address the above-mentioned problems.
When inputs are images, existing solutions for root cause analysis generate heatmaps  that use colors to capture the importance of pixels in their contribution to a DNN result~\cite{Selvaraju17,Montavon2019}. 
By inspecting the heatmaps generated for a set of erroneous results, a human operator can determine that these heatmaps highlight the same objects, which may suggest the root cause of the problem (e.g., long hair~\cite{Selvaraju17}). 
Based on the identified root cause, engineers can then retrain the DNN using additional images with similar characteristics. 
Unfortunately, this process is expensive and error-prone because it relies on the visual inspection of many generated heatmaps. 

In this paper, we present HUDD, the toolset that automates our methodology for the identification of root causes for DNN errors and DNN retraining~\cite{HUDD:TRel}. 
\APPR relies on hierarchical agglomerative clustering~\cite{King:2014} to identify the distinct root causes of DNN errors.
It configures clustering with a specific distance function based on the heatmaps computed for internal DNN layers.
A subset of the images belonging to each cluster is inspected by the engineer who is thus helped in determining unsafe conditions (i.e., commonalities among the images in a same cluster), including the infrequent ones (i.e., clusters with few members).
Further, \APPR relies on the computed clusters to identify new images to be used to retrain the DNN. 
Given a potentially large set of unlabeled images, \APPR 
selects the subset of images that are more likely to belong to the identified clusters.
These images are assumed to include potentially unsafe conditions and are then used to retrain the DNN.
We performed an empirical evaluation on six DNNs.
Our empirical results show that \APPR can automatically and accurately identify the different root causes of DNN errors.
Also, our results show that the \APPR retraining process improves DNN accuracy up to 30.24 percentage points and is more effective than baseline approaches.

In the remaining sections, we present our methodology, outline the tool, highlight the findings from our evaluation of \APPR with six case studies, and compare with related work.


\section{The \APPR methodology}
\label{sec:overview}

 \begin{figure}[tb]
 \includegraphics[width=\columnwidth]{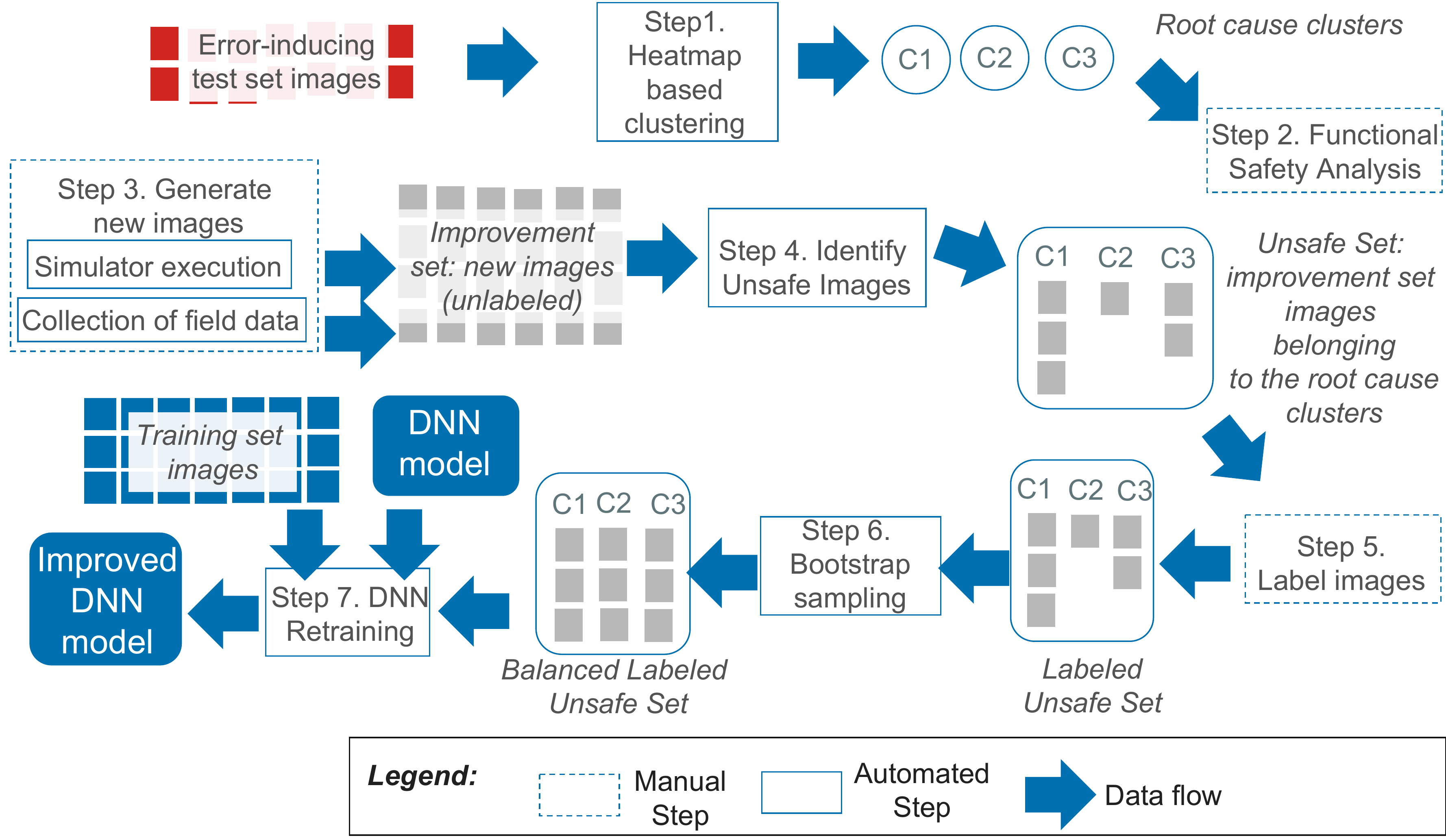}
 \caption{Overview of the \APPR methodology.}
 \label{fig:approach}
 \end{figure}

\APPR works in seven steps, depicted in Figure~\ref{fig:approach}.


In \emph{Step 1}, \APPR takes as input the test set images leading to a DNN error (hereafter, error-inducing test set images). A DNN error might be either a wrong output label (for classifier DNNs) or an output loss higher than a given threshold (for regression DNNs).  
Step 1 consists of three activities: (1) generate heatmaps for the error-inducing test set images, (2) compute a distance matrix with the distances between every pair of images\footnote{We rely on the Euclidean distance function applied to the heatmap of each image.}, and (3) execute hierarchical agglomerative clustering to group images based on the computed distances. 
Step 1 
leads to the identification of \emph{root cause clusters (RCCs)}, i.e., clusters of images with a common root cause for the observed DNN errors.


To generate heatmaps, HUDD relies on the Layer-Wise Relevance Propagation (LRP) algorithm~\cite{Montavon2019}. 
LRP enables the generation of heatmaps for the internal layers of the DNN (\emph{internal heatmaps}).
An internal heatmap for a layer $k$ consists of a matrix with the relevance scores computed for all the neurons of that layer.

\begin{figure}[tb]
\includegraphics[width=8.4cm]{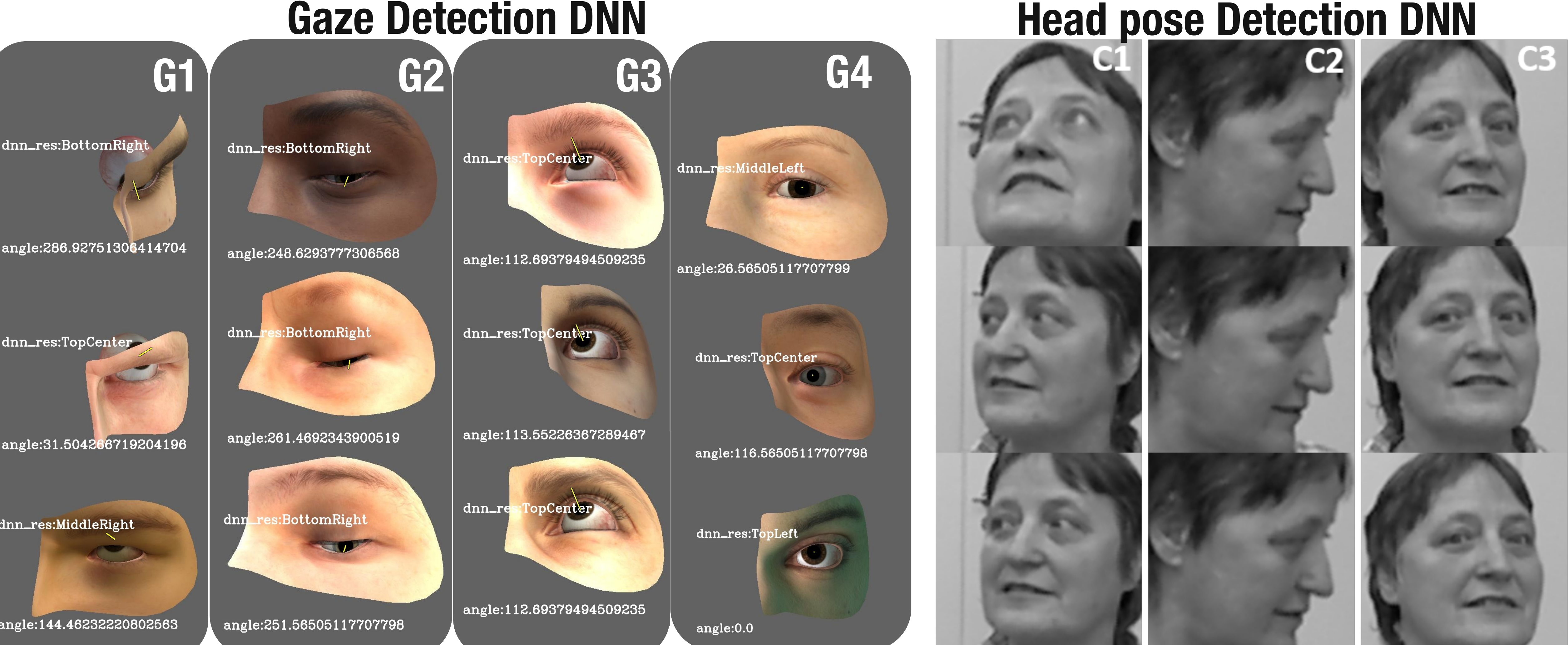}
\caption{Example of clustering results for two DNNs.}
\label{fig:ClusteringResults}
\end{figure}

In \emph{Step 2}, engineers inspect the RCCs (typically a subset of the RCC images) to identify unsafe conditions, as required by functional safety analysis. Figure~\ref{fig:ClusteringResults} provides examples of RCCs generated for a DNN that detects the gaze of an eye (Gaze-DNN) and for a DNN that determines the position of a person's head (HPD-DNN). 
Gaze-DNN processes images generated with a simulator; HPD-DNN processes real-world images.

The clusters in Figure~\ref{fig:ClusteringResults}
show that 
\APPR identifies root causes that are associated with: (1) \emph{borderline cases}
(e.g., the gaze and head pose angle detected by G3, C1, and C2), 
(2) \emph{an incomplete training set} (e.g, persons with a face turned left and eyes looking right in C3),
(3) \emph{an incomplete definition of the predicted classes}
(i.e., the middle center gaze detected by cluster G4 and the closed eyes detected by cluster G2) and (4) \emph{limitations in our capacity to control the simulator} (i.e., unlikely face positions detected by cluster G1). 
The first two cases are addressed by HUDD retraining procedures (i.e., Steps 4-7); also, borderline cases may help engineers identify countermeasures (e.g., an additional camera with a different camera angle), the identification of countermeasures being part of standard safety analysis practices. Finally, the other causes require that engineers modify the DNN (e.g., to add an output class) or improve the simulator.



In \emph{Step 3}, engineers rely on real-world data or simulation software
to generate a new set of images to retrain and improve the DNN. Step 3 is common practice and entails limited effort (e.g., acquiring field images or configuring a simulator).

In \emph{Step 4}, \APPR automatically identifies the subset of images belonging to the improvement set that are likely to lead to DNN errors, referred to as  \emph{unsafe set}. It is obtained by assigning 
the images of the improvement set to the RCCs according to their heatmap-based distance. 
Since RCCs characterize only the portion of the input space that is unsafe, images belonging to the improvement set may not belong to any of these clusters. 
For this reason, \APPR selects only images that are sufficiently close to cluster members based on test set observations. 

In \emph{Step 5}, engineers manually label the images belonging to the unsafe set. Different from traditional practice, which consists of labelling a large set of additional images and select for retraining the ones that lead to DNN errors,  \APPR requires that engineers label only a small subset of the improvement set (i.e., the images that likely lead to DNN errors because they belong to the identified RCCs). 

In \emph{Step 6}, to improve the accuracy of the DNN for every root cause observed, 
independently from their frequency of appearance,
\APPR balances the unsafe set with bootstrap resampling~\cite{ML:Book}, i.e., it randomly duplicates the images belonging to the cluster until every cluster has the same size. 

In \emph{Step 7}, \APPR retrains the DNN model by relying on a training set that consists of the union of the original training set and the balanced labeled unsafe set. HUDD uses the available model to set the initial configuration for the DNN weights. The original training set is retained to avoid reducing the accuracy of the DNN for parts of the input space that do not show any error in the test set.
The retraining process is expected to lead to an improved DNN model compared to that based on the original training set.

By driving the retraining based on the observed DNN errors, HUDD enables engineers to demonstrate that they took measures to improve safety, an important aspect to comply with regulations (e.g., ISO/PAS 21448 highlights the importance to adopt methods that reduce the unsafe portion of the input space~\cite{SOTIF}).

Steps 2, 3, and 5 are manual steps which are also part of state-of-the-art solutions. But in the case of \APPR, the manual effort required in such steps is much more limited than in existing approaches. With \APPR, in Step 2, engineers inspect a few images per root cause clusters rather than the whole set of images resulting in significant cost savings and effective guidance towards the identification of root causes. Step 3 is common practice and entails limited effort such as buying field images or configuring a simulator. Finally, in Step 5, engineers label only a subset of the improvement set containing images that are likely to be unsafe and can be effectively used for retraining. Without \APPR, engineers would have to label a randomly selected subset of the improvement set, which would contain fewer unsafe images and thus be less effective during retraining.


\section{Architecture \& Usage}
\label{sec:implementation}

 \begin{figure}[tb]
 \begin{center}
 \includegraphics[width=\columnwidth]{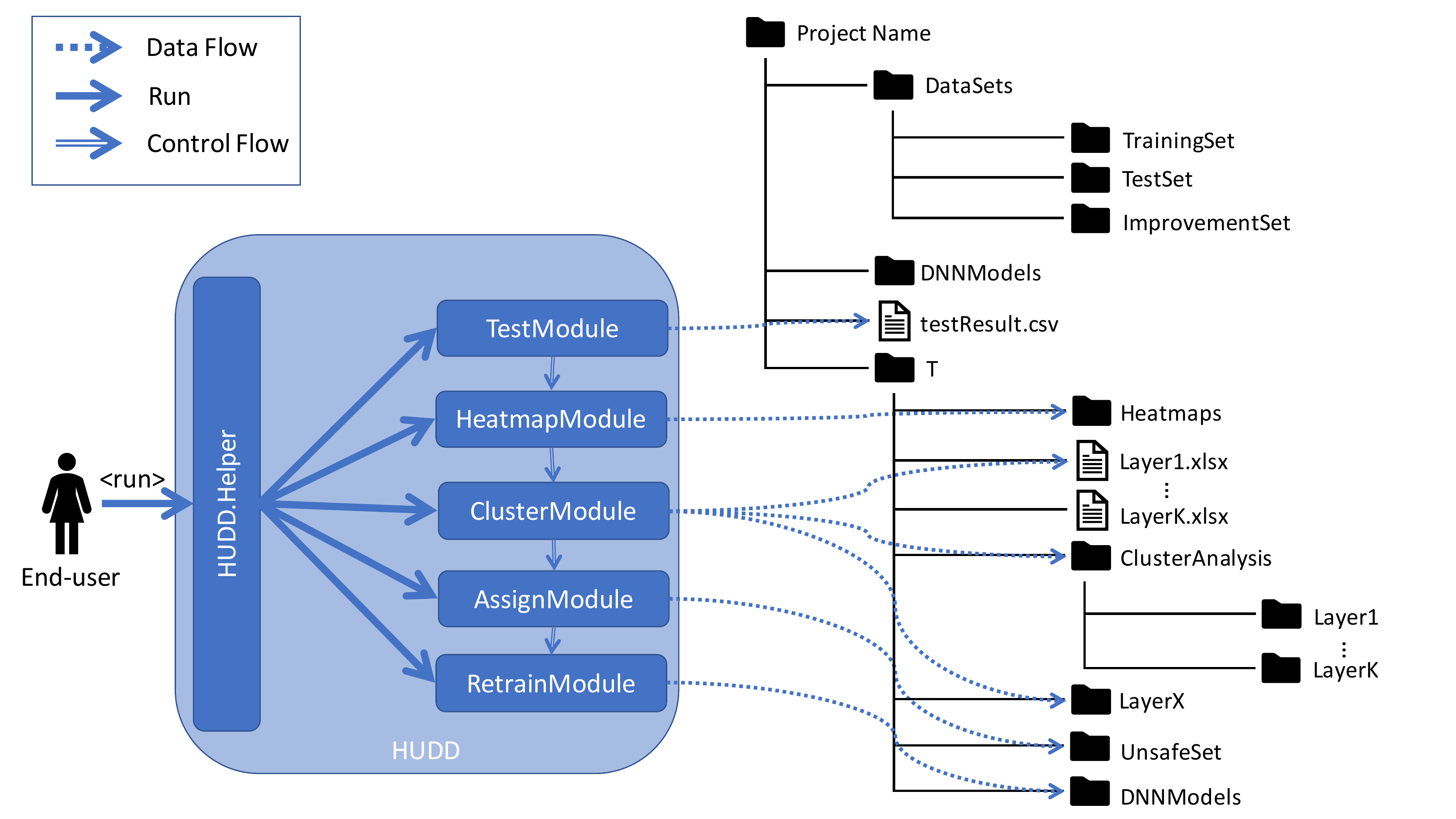}
 \caption{Architecture (left) and output structure (right).}
 \label{fig:tool}
 \end{center}
 \end{figure}

Figure~\ref{fig:tool} provides the architecture of the \APPR toolset.
\APPR is implemented in Python with the following dependencies.
\emph{Tensorflow}~\cite{tensorflow} and \emph{PyTorch}\cite{PyTorch} are used for DNN models.
\emph{NumPy}, an array programming library~\cite{numpy}, is used for the manipulation and storage of heatmaps.
\emph{Pandas}, a data analysis tool~\cite{pandas}, is used to compute heatmap distance matrices.
\emph{Scipy}, a library for scientific computing~\cite{SciPy}, is used for the generation of clusters.
\emph{Kneed}, a library that implements the kneedle algorithm~\cite{kneedle}, is used to determine the optimal number of clusters.


The HUDD tool consists of a command line user interface called \emph{HUDD.Helper} and five modules: TestModule, HeatmapModule, ClusterModule, AssignModule, RetrainModule.

To execute HUDD, 
the engineer provides to the \emph{HUDD.Helper} the DNN model to be analyzed. 
The DNN under analysis shall be stored in the \emph{DNNModels} folder; the datasets shall be provided in the \emph{DataSets} folders \emph{TrainingSet}, \emph{TestSet}, and \emph{ImprovementSet}.
\APPR relies on LRP and thus requires DNN models that integrate the LRP backpropagation function. 
We provide an AlexNet model to be used for classification tasks and a KPNet model that can be used for regression tasks. 
However, DNN models can be easily extended to integrate LRP-based heatmap generation by following existing guidelines~\cite{ExplainMontavon}.
For example, our AlexNet implementation is an extension of the PyTorch AlexNet.
The HUDD.Helper orchestrates the execution of all the other modules. The intermediate results generated by \APPR are stored within the temporary folder \emph{T}, which is kept to enable further inspection of all the processed data.

The \emph{TestModule} uses the DNN under analysis to process the inputs in the training and test set. Outputs are exported in the files \emph{trainResult.csv} and \emph{testResult.csv}. The former is used to compute training set accuracy, which is used to determine if the improved DNN is better than the original one (\APPR Step 7). The latter is used to determine which are the error-inducing images to be used to generate RCCs (\APPR Step 1).

The \emph{HeatmapModule} generates heatmaps for error-inducing images. For each DNN layer, it stores, in the \emph{Heatmaps} directory, a NumPy file with the heatmaps of all the error-inducing images.

The \emph{ClusterModule}, for each layer, computes the distance matrix and exports it in an \emph{XLSX} file. 
Also, it performs hierarchical agglomerative clustering based on the heatmaps generated for each layer and selects the optimal number of clusters. 
Finally, for each $K^{th}$ layer, it stores the generated clusters in a directory called \emph{T/ClusterAnalysis/Layer{K}}.
The clusters for the layer showing the best results (layer $X$)  are copied in the parent folder (i.e., \emph{./T/Layer{X}/}). 
For each RCC, the \emph{ClusterModule} generates a directory with all the images belonging to the cluster,
which are to be visually inspected by engineers as per HUDD Step 2.
 
To simplify the inspection of RCCs, the \emph{ClusterModule} also generates a set of animated GIF images, one for each cluster. Each generated GIF image shows all the images belonging to a cluster one after the other. Animated GIFs enable engineers to inspect a large number of images in a few seconds (e.g., we configure our tool to visualize 100 images per minute) thus facilitating the detection of the common characteristics among them.

The \emph{AssignModule} processes the ImprovementSet images and stores the unsafe set in the  folder \emph{UnsafeSet}. Finally, the \emph{RetrainModule} retrains the DNN using the images in the training and unsafe sets. 
The retrained DNN model is saved in the DNNModels directory.


Our toolset, case studies, and results are available online~\cite{REPLICABILITY}.

\section{Empirical Evaluation}
\label{sec:evaluation}

This section provides an overview of the main findings of an evaluation conducted to address the following research questions~\cite{HUDD:TRel}:
\begin{itemize}
\item[RQ1] Does \APPR enable engineers to identify the root causes of DNN errors?
\item[RQ2] How does \APPR compare to traditional DNN accuracy improvement practices?
\end{itemize}

For our empirical evaluation, we considered six DNNs. A gaze detection system (GD) that determines the gaze direction of a human face.
A drowsiness detection system (OC) that features the same architecture as the gaze detection system, except that the DNN predicts whether eyes are closed.
A head poses detection system (HPD) that classifies the position of a person's head in an image according to nine classes: straight, bottom-left, left, top-left, bottom-right, right, top-right, reclined, looking up.
A facial landmarks detection system (FLD) that identifies the location of the pixels corresponding to 27 face landmarks delimiting seven face elements.
An object detection system (OD) that tries to detect the existence of eyeglasses.
A traffic signs recognition system (TSR) that recognizes the presence of traffic signs in road images.

RQ1 investigates whether HUDD is feasible and generates RCCs with images presenting a common set of characteristics that are plausible causes of DNN errors. 

To determine if the RCCs generated by \APPR include images with common characteristics, we relied on images generated using simulators. Since simulator images can be associated with the simulator parameter values used to generate them (e.g., the vertical angle of a person’s head), we could objectively determine if the images in the same cluster present common characteristics. Indeed, a characteristic that is shared between the images in the same RCC shall lead to a lower within-cluster variance, for at least one simulator parameter, compared to the whole test set. Also, by focusing on the parameters showing a high variance reduction (e.g., >50\%), we can objectively determine if the RCCs help engineers spot the root cause of an error. Indeed, if the average value for such parameters is close to a value that likely leads to error-inducing images (e.g., a gaze angle that is borderline between two gaze directions) we can assume that the RCC provides an explanation for the DNN error that can be understood by the engineer inspecting the images.

Our results show that a very high percentage of the clusters (i.e., between 57\% and 100\%) include at least one parameter with 50\% reduction in variance, which means that, for most of the clusters, engineers can identify commonalities among images. Also, all the parameters with high reduction in variance are associated with image characteristics that are plausible causes of errors. 

To evaluate if the visual inspection of root cause clusters is practically feasible, we report on the number of clusters generated by HUDD. Precisely, the ratio of error-inducing images that should be visually inspected when relying on HUDD, should provide an indication of the time saved with respect to current practice (i.e., manual inspection of all the error-inducing images). To perform the evaluation, based on our experience, we assumed that engineers visually inspect five images for each root cause cluster. Table~\ref{table:RQ1.1} provides summary data; we can observe that the ratio of error-inducing images that is inspected with \APPR is low, ranging from 1.49\% (GD) to 22.84\% (FLD), with a median of 6.87\%. This suggests that the analysis supported by \APPR saves a great deal of effort with respect to current practice (i.e., manual inspection of all the error-inducing images). A user study concerning the time savings introduced by HUDD is part of our future work

\begin{table}[tb]
\footnotesize
\caption{Percentage of manually inspected images for each case study DNN.}

\begin{tabular}{
|@{\hspace{1pt}}p{1.2cm}
|@{\hspace{1pt}}p{20mm}
|p{15mm}
|p{25mm}|
}
\hline
\textbf{Case study DNN}
&\textbf{\# of failing images}&\textbf{\# Root cause clusters}&\textbf{Percentage of manually inspected images}
\\
\hline
GD&5371&16&1.49\%\\
OC&506&14&13.82\%\\
HPD&1580&17&5.38\%\\
FLD&1554&71&22.84\%\\
OD&838&14&8.35\%\\
TSR&2317&20&4.31\%\\
\hline
\end{tabular}\\
\label{table:RQ1.1}
\end{table}%

RQ2 concerns DNN improvement. We considered four DNNs working with simulator images and two DNNs working with real-world images. We compared HUDD with two baseline approaches: (1) retraining with failing images selected from a subset of the improvement set and (2) retraining with random images. To avoid bias, for all the considered approaches, we label the same number of images. Experiments were repeated ten times. HUDD leads to significantly larger accuracy improvements than baselines,
increasing DNN accuracy up to 30.24\%. 


\section{Related Work}
\label{sec:related}

A number of tools supporting DNN explanation are available nowadays~\cite{ExplainMontavon,huang2018survey}.
However, for explanations concerning DNNs that process images, such frameworks boil down to generating heatmaps
one for every error-inducing input image,
that shall be visually inspected by engineers. 
Example frameworks are INNvestigate~\cite{innvestigate} and TorchRay~\cite{TorchRay}.
The cost of the manual inspection of heatmaps is one of the problems addressed by HUDD.

Research on the automated debugging and repair of DNNs is still at very early stages and includes MODE~\cite{Ma2018} and Apricot\cite{Zhang19}. 
Similarly to \APPR, MODE automatically identifies the images to be used to retrain a DNN~\cite{Ma2018}. 
However, it cannot identify the root causes of DNN errors, which is a major limitation in our context.
Also MODE entails repeated modification and retraining of the DNN under test, which is an expensive endeavor. 
Further, no tool implementing MODE is available.
Finally, in our empirical evaluation, we evaluated HUDD with an Object Detection (OD) classifier DNN that has been used for the evaluation of MODE. Despite differences in the DNN architecture used by HUDD and MODE, both models are trained on the same images. While MODE improves the model’s overall accuracy from 83\% to 89\% (+6\%), HUDD improves the model’s overall accuracy from 84\% to 97\% (+13\%).
 Apricot~\cite{Zhang19} repairs DNNs by changing the weights of the DNN model; however, an implementation of Apricot is not available.
HUDD is the first tool for the automated debugging of DNNs that is available for reuse.

\section{Conclusion}
\label{sec:conclusion}

We introduced \APPR, a toolset that supports safety analysis practices for DNN-enabled safety-critical systems. 
It generates clusters (i.e., root cause clusters, RCCs) containing misclassified input images 
sharing a common set of characteristics that are plausible causes of errors. 
In addition, \APPR minimizes the effort required to select and label 
additional images to be used to augment the training set and improve the DNN.

Empirical evaluation with simulator images show that \APPR generates clusters 
with images that provide explanations for DNN errors; further, results with both simulated and real images show how these clusters can be effectively used to select new images for retraining, in a way that is more efficient than existing practices and leading to better DNN accuracy. 

\APPR is a software engineering tool to support the development of ML-based systems. Indeed, by helping identify different plausible causes of DNN errors, it supports engineers in specifying solutions to improve the system. For example, (1) RCCs that highlight an incomplete training set suggest further training, whereas (2) RCCs with borderline cases may suggest introducing technical countermeasures. Further, the automated retraining strategy implemented by \APPR, in addition to supporting automated debugging, enables engineers to justify their selection of retraining images according to safety principles (i.e., to show the intent of eliminating root causes of errors).

\balance

\begin{acks}
This project has received funding from IEE Luxembourg,
Luxembourg’s National Research Fund (FNR) under grant 
BRIDGES2020/IS\-/14711346/FUNTASY,
 the European Research Council (ERC) under the European
Union’s Horizon 2020 research and innovation programme (grant agreement No 694277), and
NSERC of Canada under the Discovery and CRC programs. 
\end{acks}

\bibliographystyle{ACM-Reference-Format}
\bibliography{./bibliography/HUDD}

\end{document}